\documentstyle[epsfig,psfig]{mn}
\normalsize

\def\Mpch{\, h^{-1}{\rm Mpc}}    
\def\kpch{\, h^{-1}{\rm kpc}}    
\def\Mpc{\,{\rm Mpc}}    
\def\kms{\, {{\rm km}}\,{{\rm s}}^{-1} }
\def\LCDM{\rm{$\Lambda$CDM}\,}
\def\geqsim{\lower.73ex\hbox{$\sim$}\llap{\raise.4ex\hbox{$>$}}$\,$}
\def\leqsim{\lower.73ex\hbox{$\sim$}\llap{\raise.4ex\hbox{$<$}}$\,$}

\begin{document}

\title
[On Statistical Lensing and the Anti-Correlation Between 2dF QSOs and Foreground Galaxies]
{
On Statistical Lensing and the Anti-Correlation Between 2dF QSOs and Foreground Galaxies
}

\author[A.~D.~Myers et al.]
{
A.~D.~Myers$^{1,6}$\footnotemark,   P.~J. Outram$^1$,  T. Shanks$^1$,  B.~J. Boyle$^{2}$, S. M. Croom$^{2}$, 
\newauthor   N.~S. Loaring$^{3}$, L. Miller$^{4}$, \& R.~J. Smith$^{5}$
\\
1 Department of Physics, Science Laboratories, South Road, Durham, DH1 3LE, U.K.
\\
2 Anglo-Australian Observatory, PO Box 296, Epping, NSW 2121, Australia\\
3 Mullard Space Science Laboratory, University College London, Holmbury St. Mary, Dorking, Surrey,
RH5 6NT, U.K.\\
4 Department of Physics, Oxford University, Keble Road, Oxford, OX1 3RH, U.K.\\
5 Liverpool John Moores University, Twelve Quays House, Egerton Wharf,
Birkenhead, CH41 1LD, U.K. \\
6 Department of Astronomy, University of Illinois, 1002 W Green Street,
Urbana, Il 61801, U.S.A.\\
}

\maketitle 
 
\begin{abstract}

We study the cross-correlation of APM and SDSS galaxies with  
background QSOs taken from the  2dF QSO Redshift Survey, and detect
a significant  ($2.8\sigma$) anti-correlation. The lack of a similar signal between
stars in the 2dF survey and our galaxy samples gives us confidence that
the anti-correlation is not due to a
systematic error.  The possibility that  dust in the foreground galaxies could produce such
an anti-correlation is marginally rejected, at the $2\sigma$ level through consideration of the colours of
QSOs behind these galaxies.  It is possible that a dust model that
obscures QSOs without reddening them, or preferentially discards
reddened QSOs from the 2QZ sample, could produce such an
anti-correlation, however, such models are at odds with the positive
QSO-galaxy correlations found at bright magnitudes by other authors.

Our detection of a galaxy-QSO anti-correlation is consistent with
the predictions of statistical lensing theory.  When combined 
with earlier results that have reported  a {\it positive} galaxy-QSO correlation,  a consistent and compelling picture
emerges that spans faint and bright QSO samples showing positive or
negative correlations according to the QSO $N(m)$ slope.

We find that galaxies are highly anti-biased on small
scales.  We consider two models that use quite
different descriptions of the lensing matter and find they yield consistent
predictions for the strength of galaxy bias on $0.1\Mpch$ scales of $b
\sim 0.1$ (for a \LCDM cosmology).   Whilst the slope 
of our power-law fit to the QSO-galaxy cross-correlation does not
allow us to rule out a linear bias parameter, when we compare our
measurement of $b$ on $100\kpch$ scales to independent methods that
determine $b \sim 1$ on $\Mpch$ scales, we conclude that bias, on these
small scales, is scale-dependent. These results indicate
that there appears to be more mass present, at least on the $100\kpch$ scales probed, than
predicted by simple
\LCDM biasing prescriptions, and thus constrains halo
occupation models of the galaxy distribution.

\end{abstract}

\begin{keywords}
{surveys - quasars, quasars: general, large-scale structure of
Universe, cosmology: observations, gravitational lensing}
\end{keywords}

\footnotetext{E-mail: adm@astro.uiuc.edu}

\section{Introduction}
\label{sec:intro5}
Intervening massive structures can, via weak gravitational
lensing, alter the density of high-redshift objects detected
behind them. 
Myers et al. (2003) demonstrated the anti-correlation between
faint QSOs and groups of galaxies, confirming the earlier detection by Boyle,
Fong \& Shanks (1988).
Using models of lensing by simple haloes (Croom \& Shanks 1999) 
they concluded that, if due to gravitational
lensing, the observed anti-correlation 
favoured more mass in groups of
galaxies than accounted for in a universe with density parameter 
$\Omega_m = 0.3$. 

Though statistical lensing is most
apparent for larger concentrations of mass, such as groups of galaxies, most recent
attempts to measure
and model associations between QSOs and foreground mass have focussed
on the cross-correlation between QSOs and individual galaxies.
Seldner \& Peebles (1979) were among the first authors to record a 
statistically significant clustering of individual galaxies with bright, 
high-redshift QSOs, although Tyson (1986) appears to be the first 
to have mentioned lensing as a possible explanation.  Webster et al. (1988) developed a statistical 
lensing explanation for the association of QSOs with 
foreground galaxies, suggesting that more lensing mass must be being 
traced than expected \cite{Kov89,Nar89,Sch89}. Since then, more authors have found 
positive correlations between optically-selected QSOs and 
galaxies \cite{Tho95,Wil98}. Many more have measured 
positive correlations between galaxies and radio-selected 
QSOs \cite{Fug90,Bar97,Ben97,Nor00}, for which a larger lensing effect is expected.
Very few
{\it anti}-correlations between QSOs and galaxies have been detected \cite{Ben97,Fer97}, and these were mainly attributed to QSO-selection effects. However, as detected by Myers et al. (2003), an anti-correlation between
{\it faint} QSOs and foreground matter is predicted by statistical
lensing.

The galaxy distribution, relative to underlying mass, is a function of bias, whereas
the QSO distribution directly traces the mass that lensed it.  Thus, comparing the QSO-galaxy cross-correlation to the galaxy auto-correlation can be used to constrain galaxy bias. Recent observations of positive
correlations between bright QSOs and galaxies \cite{Wil98,Gaz03} suggest a very low
value for the bias parameter ($b$~$\sim$~0.1).   The work of Gazta\~naga (2003) suggests $b$~$\sim$~0.1 on sub-Megaparsec scales.  The work of Williams \& Irwin (1998) suggests $b$~$\sim$~0.1 on Megaparsec scales but is consistent with $b$~$\sim$~0.1 on sub-Megaparsec scales.  However, measurements of the strength of bias from clustering in galaxy 
surveys out to (redshift) $z \sim 0.2$ \cite{Ver02}, comparisons of local galaxy clustering with the Cosmic 
Microwave Background \cite{Lah02} and weak lensing
shear \cite{Hoe02} at $z \sim 0.35$, seem 
to converge on a linear model of bias with $b$, of the order of
unity on scales of $5-100 \Mpc$.  Taken together, these results suggest
that either there is a strong scale-dependence to bias, or that there exists an unexpected,
strong systematic effect inducing positive correlations between QSOs
and foreground galaxies.

The faint flux limit of the 2dF QSO Redshift Survey \cite{Cro04},
henceforth referred to as the `2QZ', may test these two possibilities. As the
2QZ number-magnitude count slope at the survey flux limit is relatively flat, statistical
lensing predicts an anti-correlation between 2QZ QSOs and foreground
galaxies, as opposed to the positive correlations predicted for
brighter QSO samples and detected by Williams
\& Irwin (1998) and Gazta\~naga (2003). As there is no other
explanation for such opposite signals in the different samples, such a
detection would provide a strong confirmation of the lensing
hypothesis, and hence the constraints on $b$.  There are additional reasons why the 2QZ is appropriate to study statistical lensing.  The 2QZ contains around half as many
confirmed stars as QSOs, which may be used as
a control, to determine if any 
cross-correlation arises from target-selection. The 2QZ can limit the most
likely systematic, intervening dust---by considering QSO
colours, the extent that dust in galaxies could obscure background QSOs may be determined.  Finally, the known redshifts of 2QZ QSOs can ensure that they are physically removed from a galaxy sample.  

This paper regards the cross-correlation between faint QSOs and 
foreground galaxies, and 
its implications for cosmological parameters, particularly for galaxy 
bias.  In Section~\ref{sec:data5} 
we outline the samples of QSOs and galaxies we shall cross-correlate.  In
Section~\ref{sec:cross5} we outline our cross-correlation methodology, and investigate
possible explanations for the resulting signal.
Section~\ref{sec:model5} introduces models we use to investigate
the cross-correlation of QSOs and galaxies in terms of statistical
lensing. Section~\ref{sec:discuss5} applies these lensing models and discusses implications for cosmological parameters,
especially how galaxies are biased relative to underlying
matter. Finally, in Section~\ref{sec:summary5}, we summarise the main
results of this paper.

\section{QSO and Galaxy Samples}
\label{sec:data5}

\begin{figure}
\centering
\includegraphics[width=8cm]{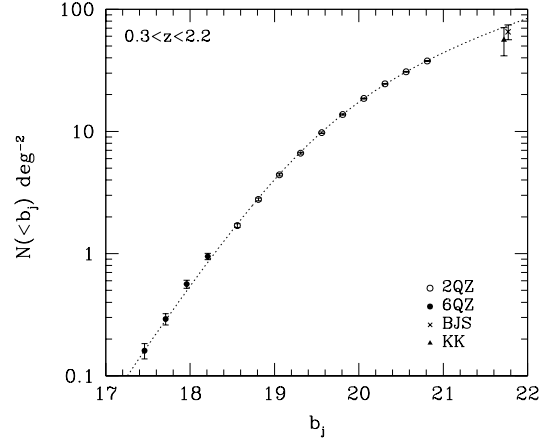}
\caption{\small{
The QSO integrated number counts, for the
  2QZ, in 0.2~mag bins, with Poisson errors. The line is a smoothed power law fit to
  the differential number counts.  Brighter data points are from the
  6QZ. Also displayed are the faint data 
  from Boyle, Jones \& Shanks (1991) and Koo \& Kron (1988), which have 
  been offset slightly to prevent the points from merging.}}
\label{fig:Nmag.eps}
\end{figure}

The QSO and galaxy samples we will cross-correlate are essentially the 
same as described by Myers et al. (2003), so we will only 
outline them briefly. QSOs are taken from the final 2QZ catalogue 
\cite{Cro04}.  The 2QZ comprises two $5\deg \times 75\deg$
declination strips, one in an
equatorial region in the North Galactic Cap (centred at $\delta=
0^{\circ}$ with 09$^{\rm h}$50$^{\rm m}$ $\la \alpha \la $ 14$^{\rm
  h}$50$^{\rm m}$), and one at the South Galactic Pole (centred at $\delta = -30^{\circ}$, with 21$^{\rm h}$40$^{\rm m}$ 
$\la \alpha \la $ 03$^{\rm h}$15$^{\rm m}$).
We will refer to these regions as the `NGC' and `SGC' respectively.
QSOs are selected by
ultra-violet excess (UVX) in the $u-b_{\rm J}:b_{\rm J}-r$
plane, in the magnitude range $18.25 \leq b_{\rm J} \leq 20.85$. The 2QZ colour selection is $\ga90$ per cent complete for UVX QSOs over the redshift
range $0.3<z<2.2$. At higher redshifts the UVX technique fails as the Lyman-alpha forest enters the $U$ band, and the completeness of the survey rapidly drops.
Unless otherwise specified, we will consider only QSOs with 
a redshift $z>0.4$, to prevent the overlap in real space of QSO and 
galaxy samples.  We shall work with only the most definitively 
identified QSO sample, the so-called quality `11' sample.  Quality `11' denotes a sample with the best level of reliability for both the QSO identification and redshift estimate (see Croom et al. 2004 for further explanation).  These restrictions in redshift and 
spectroscopic quality leave 12042 QSOs in the SGC and 
9565 in the NGC.  We 
shall also consider 
the supplemental 6dF QSO survey (henceforth 6QZ), which contains 376
QSOs after the application of  our $z>0.4$ and `11' only
spectroscopic identification criteria.  For further details of the 2QZ
and 6QZ, see Croom et al. (2004).

The expected strength of lensing-induced correlations between galaxies and a magnitude-limited
sample of QSOs depends on the slope of the integrated number-magnitude
counts, $\alpha$, fainter than the QSO sample's limit \cite{Nar89}.  
 An enhancement of QSOs is expected behind
foreground lenses when the slope of the QSO number-magnitude 
counts is greater than 0.4 and a deficit when the slope is less than 0.4. We thus need
to estimate this slope for any lensing analysis. In
Fig.~\ref{fig:Nmag.eps} we reproduce the QSO $N(<m)$ for the 2QZ from Myers et al. (2003).  The points are QSO number counts in 0.2~mag
  bins with Poisson errors, and have been corrected for incompleteness and absorption by dust in our Galaxy.  The line is a smoothed power law fit to the differential counts of form
  
\begin{eqnarray}
\frac{\rm d \it N}{\rm d \it m} = \frac{N_0}{10^{-\alpha_{\rm{d}}(m-m_0)} + 10^{-\beta_{
\rm{d}}(m-m_0)}}
\label{eq:slopefit}
\end{eqnarray}
  
with a steep bright-end slope ($\beta_{\rm{d}} = 0.98$), a knee at $m_0 
= 19.1$, and a flatter faint end slope ($\alpha_{\rm{d}} = 0.15$). Myers et al. (2003) marginalised across all of the parameters in this fit and determined a faint end slope of $\alpha =0.29\pm0.015$, noting that incompleteness corrections might raise the $1\sigma$ error to as much as 0.05.  In this paper, we will
assume a slope of $\alpha =0.29\pm0.03$ at the survey flux limit.  Brighter data points are from the 6QZ. Also displayed are the faint data 
  from Boyle, Jones \& Shanks (1991) and Koo \& Kron (1988), which have 
  been offset slightly to prevent points from merging.
Recent observations, attributed to statistical lensing, of positive
correlations between QSOs and galaxies \cite{Wil98,Gaz03} have used
relatively bright QSO samples, which only probe the steep QSO number-magnitude 
counts below the knee. The large size and faint flux limit of the 2QZ
allows us to probe significantly beyond the knee and test the
lensing prediction of an anti-correlation between
{\it faint} QSOs and galaxies.

The southern galaxy sample we consider in this paper is taken from 
the APM Survey \cite{Mad90b}, which is considered photometrically complete 
to a magnitude of $b_{\rm J}<20.5$ \cite{Mad90a}.  The northern galaxy sample is 
taken from the Sloan Digital Sky Survey (henceforth SDSS) Early Data 
Release (henceforth EDR) of June 2001 \cite{Sto02}.  The SDSS EDR sample 
is transformed into the $b_{\rm J}$ band from the SDSS $g'$ and $r'$ bands using the 
colour equations of Yasuda et al. (2001) and cut to $b_{\rm J}<20.5$ to match the APM limit. Both galaxy samples are 
restricted to areas in which they overlap the 2QZ strips. This leaves 
nearly 200,000 galaxies in the SGC 2QZ strip and 100,000 in the NGC 2QZ strip.  
Note that the SDSS EDR only partially fills the 2QZ NGC strip.

\section{QSO and Galaxy Cross-Correlation Functions}
\label{sec:cross5}
 
Correlation functions  \cite{Pee80} are the main statistic of 
choice in studies of how QSO and galaxy distributions are related,
although how the statistic is estimated varies considerably.  In this
section, we measure the two-point cross-correlation between SDSS or APM galaxies
and 2QZ (or 6QZ) QSOs.  We study possible selection effects and different explanations for the signal.
Notably, the expected variation of the QSO-galaxy cross-correlation with
redshift and (especially) magnitude of the QSO sample is a strong
prediction of the statistical lensing hypothesis, and is something we may be
able to test using 2QZ data.

\subsection{Correlation Estimator and Errors}

To measure the two-point correlation function, $\omega(\theta)$, we use the estimator
\cite{Pee80}
\begin{equation}
\omega(\theta) = \frac{DD_{12}(\theta)\bar{n}}{DR_{12}(\theta)} - 1,
\label{equation:corrfunc5}
\end{equation}
\noindent where $DD_{12}$ denotes the number of data-point
{\it pairs} drawn from populations 1 and 2 respectively with
separation $\theta$. For  $DR_{12}$ the population 2 data are replaced with a
catalogue of random points with the same angular selection function as the data.  The factor $\bar{n}$ is the ratio of the size of
the random catalogue to the data.
Throughout this paper, we produce random catalogues with $\bar{n}=10$, minimizing statistical noise but allowing efficient speed of calculation.  For further details on the 2QZ angular selection function see Outram et
al. (2003) and Croom et al. (2004).
Note that the cross-correlation function is equivalent in either
`direction' (i.e. under the exchange of labels $1$ and $2$ in
Equation~\ref{equation:corrfunc5}) {\it provided that the angular
  selection of sources is accounted for by an appropriate random
  catalogue}.  In general, the angular selection functions of, say, a galaxy
population and a QSO population are not the same.  Further, the angular 
completeness of a given survey is generally a function of redshift
and magnitude, so considerable care should be taken in constructing
random catalogues for population subsamples.

Numerous estimates of error on the cross-correlation have been
proposed.  We will consider three of these.  One of the simpler forms,
is the Poisson error based on the number of data-data pairs (across the entire survey) in the
angular bin probed:
\begin{equation}
\sigma_{\omega}^2(\theta) = \frac{\left[1+\omega(\theta)\right]^2}{DD(\theta)}
\label{equation:poisDD}
\end{equation}
\noindent where we will use $\sigma_{\omega}$ to denote the standard error on the 
correlation function. Given that we wish to measure the
significance of the correlation function compared to the null
hypothesis represented by the random catalogue,
we might instead use the Poisson error based on the number of data-random pairs, which could be achieved by substituting $DR/\bar{n}$ for $DD$ in Equation~\ref{equation:poisDD}.

The hypothesis that error on the correlation function is Poisson is not
strictly fair.  All else being equal, the number of counts could
be highly correlated as the same points appear in different pairs that are included in many different
bins, especially on large scales.  Some authors \cite{Sha94,Cro96} have suggested corrections to
the Poisson form of error.  Instead of such corrections, we
will consider errors from field-to-field variations in the
correlation function (see, e.g., Stevenson et al. 1985).  Our data samples 
are split into 30 subsamples.  This is arbitrary in the case of the SDSS
EDR but deliberately reflects plate boundaries in the case of the APM data (and by extension the 2QZ, which is derived from APM
photometry).  The cross-correlation function is measured for each subsample
and the variance between the subsamples is determined.  The standard
error on $\omega(\theta)$ is then the standard error
between the subsamples, inverse weighted by variance to account for different 
numbers of objects in each subsample
\begin{equation}
\sigma_{\omega}^2(\theta) =
\frac{1}{N-1}\sum_{L=1}^{N}\frac{DR_L(\theta)}{DR(\theta)}[\omega_L(\theta)
- \omega(\theta)]^2.
\label{equation:FTFerr}
\end{equation}
\noindent The weighting by number of objects is
essential, mainly as plates in the southern APM immediately East of
00h RA cover less area than other plates at the same 
declination but also because of varying completeness in the QSO catalogue.  We will refer to this error as {\it field-to-field error}.

We will also consider a similar estimate of error to
Equation~\ref{equation:FTFerr}, essentially a weighted form of 
the estimate
proposed by Scranton et al. (2002) when calculating the auto-correlation of
galaxies in the SDSS EDR
\begin{equation}
\sigma_{\omega}^2(\theta) =
\sum_{L'=1}^{N}\frac{DR_{L'}(\theta)}{DR(\theta)}[\omega_{L'}(\theta)
- \omega(\theta)]^2
\label{equation:JACKerr}
\end{equation}
\noindent  Note that $L$ (in Equation~\ref{equation:FTFerr})
refers to a subsample on one of our 30 plates, whereas $L'$ refers to the 
subsample remaining {\it on the other 29 plates}.  In other words,
the procedure outlined in Equation~\ref{equation:JACKerr} is to remove
each of the plates in turn and to calculate the variance
between each sample on the 29 remaining plates.  We shall refer to this 
as {\it jackknife error}. The unweighted version of this estimate agrees well with simulations (see the appendix
of Zehavi et al. 2002).

Finally, we will measure the statistical correlation (not to be confused with the correlation {\it function}) between adjacent bins of $w(\theta)$.  The 
statistical correlation is related to the covariance, and is essentially an estimate of how independent the bins are - 
whether bin $i$ has a tendency to take the same value as bin $j$.  We 
might expect the covariance of the correlation function to be high, as 
the same data points can appear in different pairs that are counted in many different bins.  The statistical correlation takes the following form
\begin{equation}
Corr(i,j) =  
\frac{Cov(i,j)}{\sigma(\theta_i)\sigma(\theta_j)}
\label{equation:covar}
\end{equation}
where the covariance, $Cov(i,j)$, is defined as
\begin{equation}
Cov(i,j) =  
\frac{1}{N}\sum_{M=1}^{N}(\omega_M(\theta_i) - 
\bar{\omega}(\theta_i))(\omega_M(\theta_j) - 
\bar{\omega}(\theta_j))
\end{equation}
\noindent and $\theta_i$ and $\theta_j$ are two bins at different scales, and 
$\bar{\omega}$ and $\sigma$ represent the mean and standard deviation over 
a number of realisations, $M$.  The correlation is $0$ if the 
bins are independent, approaches $1$ if an increase in bin $i$ leads to an 
increase in bin $j$ and approaches $-1$ if an increase in bin $i$ leads to 
a decrease in bin $j$.

\begin{figure}
\centering
\includegraphics[width=8cm,totalheight=8cm]{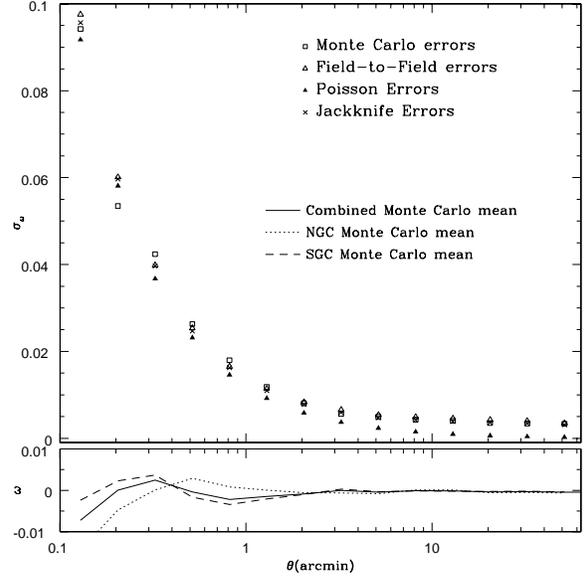}
\caption[\small{Error Estimates and Estimator Check.}]{In the upper
  panel, we compare the Poisson, field-to-field and jackknife error
  estimates with a Monte Carlo error
  estimate determined from 100
  Monte Carlo simulations of the NGC and SGC strips. In the lower
  panel we plot the mean of the 100 simulations for the NGC and SGC
  individually and for the NGC and SGC combined.}
\label{fig:sigma5}
\end{figure}

\begin{figure}
\centering
\includegraphics[width=8cm,totalheight=8cm]{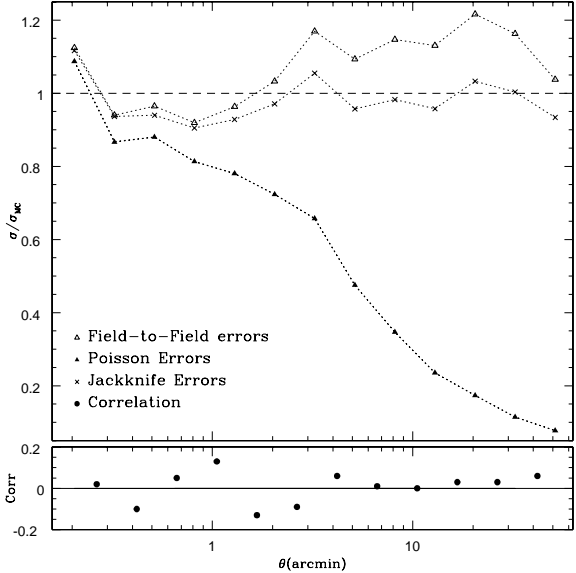}
\caption[\small{Error Estimates Compared to Monte Carlo Estimates.}]{In 
the upper 
panel we
  display the error on the correlation function taken in ratio
  to the Monte Carlo error estimate (1 standard deviation across the Monte Carlo simulations), for each estimate of error
  mentioned in the text.  In all cases the errors are determined for
  the combined NGC and SGC sample.  A dashed line is drawn for comparison at
  $\sigma$/$\sigma_{MC} = 1$ (where the
  Monte Carlo estimate itself would lie).  The 
lower panel depicts the covariance between adjacent bins determined from 
100 Monte Carlo realisations.} 
\label{fig:sigmacomp5}
\end{figure}

To test the accuracy of our correlation function estimator and the
associated error, we have created 100 Monte Carlo simulations of the 2QZ.  Each simulation contains the same number of ($z>0.4$, 
identification quality of `11') QSOs as the 2QZ and has the same angular
selection function, neglecting any intrinsic QSO clustering.  We cross-correlate the
simulated samples against APM galaxies (for the SGC strip)
and the SDSS EDR (for the NGC strip).  For each sample, the estimates
of error outlined in this section are calculated and averaged.  The mean value of the cross-correlation across the
100 samples is taken and the standard deviation ($1\sigma$) is recorded
as the Monte Carlo error.  To avoid confusion between our `Monte Carlo error' and the statistical `standard error' we shall refer to the Monte Carlo error as the `Monte Carlo deviation'.

In the lower panel of Fig.~\ref{fig:sigma5}, we display the mean
cross-correlation signal across the 100 Monte Carlo simulations.  The agreement between the NGC and SGC results are
excellent - better than 12~per~cent of the Monte Carlo deviation on the NGC 
mean over all scales.  The deviation of the combined result from zero,
the expected result, is similarly no more than
12~per~cent of the combined Monte Carlo deviation over all scales.  We note 
that the shot noise (i.e. the standard error) would comprise 10~per~cent of the Monte Carlo deviation 
(as we have 100 samples).  Although the correlation function diverges on scales smaller than 0.3~arcmin, the error is
sufficiently large on these scales that we may consider the correlation 
estimator probably valid on all of the scales plotted and certainly valid on
scales larger than 0.3~arcmin.  Note that the consistency of the
correlation estimator across all scales indicates that the software we
use to calculate the estimator is robust. Also, note that the average Monte Carlo sampling of $\omega$ contains 1180 random points in the smallest bin and has a jackknife error of 0.098---this means that the typical fluctuation in our random catalog is three-and-a-half times smaller than our quoted error, and is dwarfed by the jackknife error, justifying our choice of $\bar{n}=10$ in Equation~\ref{equation:corrfunc5}.

In the upper panel of Fig.~\ref{fig:sigma5} we compare the various error
estimates.   The general trend of the errors is in good
agreement, although the Poisson error estimate begins to under-predict
the error (as compared to the Monte Carlo estimate) on
larger scales.  We assume that the Monte Carlo deviation represents a fair
estimate of the true error on the correlation function and, In Fig.~\ref{fig:sigmacomp5}, plot the various
errors taken in ratio to the Monte Carlo deviation.  It is obvious that the 
Poisson error is an underestimate on $\ga5$~arcmin scales.  We have also estimated the Poisson error by substituting $DR$ for $DD$ in Equation~\ref{equation:poisDD}.  Such a change has negligible effect, as the simulated QSOs have a largely random distribution. The jackknife and field-to-field error estimates are much
better than the Poisson estimate and constitute reasonable estimates on scales
from 0.2~arcmin to nearly a degree.  The field-to-field error,
however, is perhaps a 20~per~cent overestimate on larger scales, where
the jackknife error remains in line with the Monte Carlo estimate. The jackknife error estimate has an additional advantage over 
the field-to-field estimate.  When the number of data points on 
any plate approaches zero, the field-to-field estimate is 
ill-defined but the jackknife estimate remains accurate until 
the number of data points across the entire survey approaches 
zero.  

In Fig.~\ref{fig:sigmacomp5}, we also plot the correlation due to the covariance 
between {\it adjacent} bins.  This covariance is low 
- almost within the 10~per~cent expected standard error.  However, this covariance measure demonstrates that no correlation between adjacent bins is artificially introduced by our random catalogues -  it does not guarantee that the data will show no covariance.  We will quote the significance of results by 
estimating the correlation function and its associated error in one 
`large' bin (usually out to 10~arcmin), to minimise any covariance in the data on these scales. Throughout this paper, we adopt the estimate of the correlation function defined by 
Equation~\ref{equation:corrfunc5},
together with the jackknife error estimate of
Equation~\ref{equation:JACKerr} as both seem fair over scales we will probe.

\subsection{The Cross-Correlation of 2QZ QSOs and Galaxies}

\begin{figure}
\centering
\includegraphics[width=8cm,totalheight=8cm]{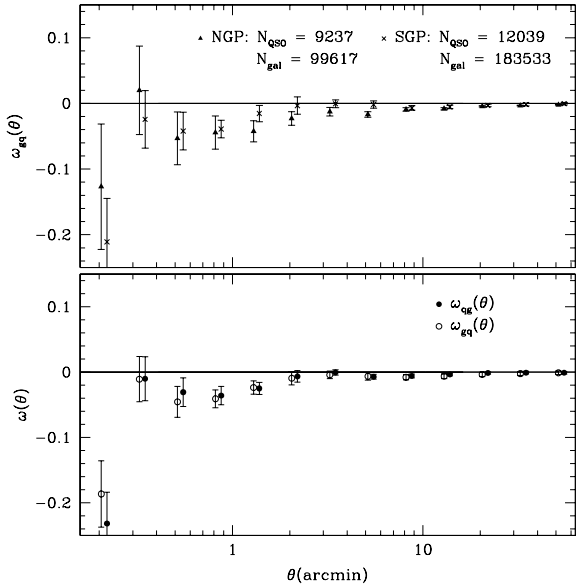}
\caption[\small{The Cross-Correlation of 2QZ QSOs and Galaxies.}]{The cross-correlation of 2QZ QSOs against SDSS EDR galaxies (in the NGC 2QZ strip) and APM galaxies (in the SGC 2QZ strip).  The upper panel displays the cross-correlation signal for the 2 strips individually.  The lower
  panel displays the signal combined for both strips.  The lower panel
  shows estimates for both `directions', centring on QSOs and
  counting galaxies ($\omega_{qg}$) and centring on galaxies and
  counting QSOs ($\omega_{gq}$).  Error bars represent $1\sigma$ 
jackknife errors.  Labels note the number of objects of each population 
present within the confines of the 2QZ boundaries. Points within the same 
bin have been offset slightly for ease of display.}
\label{fig:qsoall5}
\end{figure}

In Fig.~\ref{fig:qsoall5} we plot the cross-correlation of all 2QZ QSOs 
that meet our selection criteria ($z > 0.4$ and 2QZ
identification of `11') against SDSS EDR galaxies (in the
2QZ NGC strip) and APM galaxies (in the 2QZ SGC strip).
The upper panel of Fig.~\ref{fig:qsoall5} shows the
cross-correlation individually for the strips---galaxies
are anti-correlated with QSOs in both.  The anti-correlation is
slightly stronger in the NGC strip but not significantly so.  In the
lower panel of Fig.~\ref{fig:qsoall5}, we plot the cross-correlation
for both `directions'.  A significant anti-correlation is detected
irrespective of whether we centre on galaxies
and count QSOs or centre on QSOs and count galaxies, indicating that our 
random catalogues consistently account for the angular
selection of QSOs or galaxies.  

If we bin the data displayed in  Fig.~\ref{fig:qsoall5} in a single bin
of extent 10~arcmin and estimate the 
correlation function and ($1\sigma$) jackknife error, we find there is
an anti-correlation of strength $\omega(<10~arcmin) = -0.007\pm0.0025$  (10~arcmin is about 1 $\Mpch$ at the median galaxy redshift of 0.15).  The significance of this result is  
2.8$\sigma$ for $\omega_{qg}$ and 2.2$\sigma$ for $\omega_{gq}$.  
Although the anti-correlation is slightly less significant for 
$\omega_{gq}$, it is also slightly stronger, i.e. $\omega(<10~arcmin) = -0.008\pm0.0035$. Further, particularly on large scales ($>8$arcmin), the error in $\omega_{gq}$ is 60-80~per~cent of the error in $\omega_{qg}$.  It is unclear exactly why the errors are slightly larger when the analysis is carried 
out centring on galaxies but it is likely due to small discrepancies, perhaps second-order gradients, between the random catalogues used to mimic the different angular selection functions of the QSO and galaxy samples.

The anti-correlation displayed in 
Fig.~\ref{fig:qsoall5} is very strong on small scales and agreement 
between the two `directions' of the correlation function is excellent.
For instance, both $\omega_{gq}$ and $\omega_{qg}$ show an 
anti-correlation of strength -0.02 out to 3~arcmin at 3$\sigma$ 
significance.  Note that the innermost bin barely contributes to this 
particular signal, as it contains less than 100th of the pairs in the bin 
at 2~arcmin.  When modelling, we will consider $\omega_{qg}$, 
the slightly weaker, slightly more significant result.  This is mainly 
because ultimately, when discussing the anti-correlation in terms of 
lensing, we will compare the galaxy-galaxy auto-correlation, 
$\omega_{gg}$, to $\omega_{qg}$, which shares the same random catalogue.  
It is useful, though, to have shown a significant, consistent
anti-correlation between QSOs and galaxies irrespective of the
`direction' of the cross-correlation.

\subsection{Is the Anti-Correlation Between Galaxies and QSOs a Selection 
Effect?}
\label{sec:stars5}

Certainly, there is a significant anti-correlation between galaxies and 2QZ
QSOs.  It is natural to ask if the signal arises when constructing the QSO or galaxy catalogues.  
There are several techniques that might produce an 
anti-correlation between QSOs and galaxies.  The initial construction of 
the 2QZ UVX target catalogue removed extended images.  Although all `high' redshift ($z~\geqsim~0.5$)
QSOs should appear stellar, they may merge with foreground 
objects to look extended on images.  A bright ($b_j < 19.5$) subsample of these 
extended images should end up in the 2dF Galaxy Redshift Survey 
(henceforth 2dFGRS) and thus appear in 
deficit in the 2QZ.  It turns out that 
this affects separations between galaxies and QSOs on scales 
of about 8~arcsec \cite{Mad02}, smaller than the scales we are 
probing.  Restrictions on the placement of 2dF fibres means the 
minimum angle between objects in a 2QZ field is about 30~arcsec,  
which might mean a paucity of objects at small separations.  
However, this restriction should not include QSO-galaxy separations as, although the 2QZ was 
carried out simultaneously with the 2dFGRS, QSO observations were given
a higher observational priority, so we would expect few QSOs to be
rejected because of their proximity to galaxies.  Additionally, the vast majority of
fields in the 2dF survey were repeatedly observed to circumvent fibre-allocation 
problems.  Myers et al. (2003) suggest that a 30~arcsec restriction on the minimum angle between 2dF objects in any field leaves no significant signature on these scales.

\begin{figure}
\centering
\includegraphics[width=8cm,totalheight=8cm]{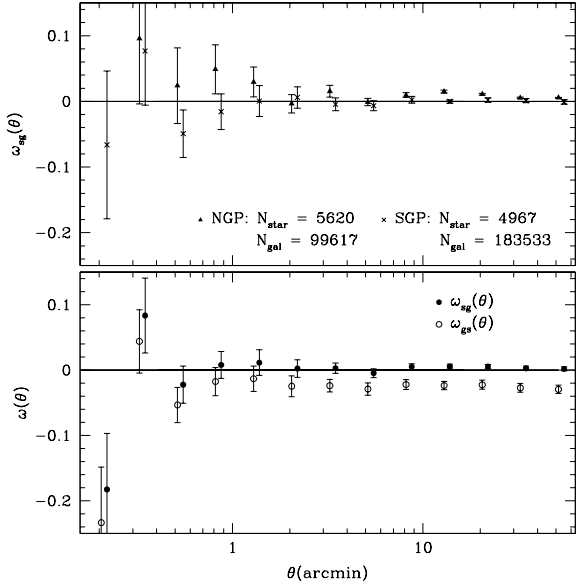}
\caption[\small{The Cross-Correlation of 2QZ Stars and Galaxies.}]{The cross-correlation of 2QZ stars against
  SDSS EDR galaxies (in the NGC 2QZ strip) and against APM galaxies
  (in the SGC 2QZ strip).  The upper panel displays the
  cross-correlation signal for the 2 strips individually.  The lower
  panel displays the signal combined for both strips.  The lower panel
  shows estimates for both `directions', centring on stars and
  counting galaxies ($\omega_{sg}$) and centring on galaxies and
  counting stars ($\omega_{gs}$).  Error bars represent $1\sigma$ 
jackknife errors. Labels note the number of objects of each population
present within the confines of the 2QZ boundaries.  Points within the 
same
bin have been offset slightly for ease of display.}
\label{fig:starall5}
\end{figure}

The easiest way to judge measurement systematics in the 2QZ is to consider 
a control sample of objects that underwent identical data reduction as the 
QSOs but should display no cosmological signatures.  There are 10587 stars 
(with `11' identification quality) in the 2QZ.  In 
Fig.~\ref{fig:starall5} we plot the cross-correlation of these stars 
against our galaxy samples.  The upper panel of 
Fig.~\ref{fig:starall5} compares the 
cross-correlation estimate for the NGC and SGC 2QZ strips.  The agreement 
is reasonable, although the NGC sample is slightly more positively 
correlated with galaxies.  Note that we might not expect 
the stellar correlation functions to be zero on all scales---gradients exist in local structure, which are not recreated in the random catalogues and which will make the correlation signal higher or lower on 
average.  However, we would expect the stellar signature to be 
{\it flat} across the scales of interest.  This is highlighted in the 
lower panel of Fig.~\ref{fig:starall5}, where we display the star-galaxy 
and galaxy-star cross correlations.  Unlike in the QSO-galaxy case, there 
is a discrepancy in the large-scale zero-point of the correlation function  
that depends upon the `direction' of the cross-correlation.  As we might 
expect, when the random catalogue is constructed to match the stellar 
distribution, the zero-point of the correlation function drops 
significantly (to about -0.03).  The large-scale value of the correlation 
function {\it is} zero when the random catalogue is constructed to 
match the galaxy distribution, which is free from (at least genuine 
physical) gradients.  The key point, is that the correlation functions 
for stars and galaxies are flat (deviating 
at most 0.5$\sigma$ from their respective zero-points on scales larger 
than $\theta\sim0.4$~arcmin), indicating that systematics in the 
construction of the galaxy and QSO samples are low, hence induce 
no false correlations in our QSO-galaxy 
cross-correlations.  The innermost points 
($\theta<0.4$~arcmin) plotted in 
Fig.~\ref{fig:starall5} seem to genuinely deviate from the 
zero-point and may be representative of merged images, fibre placement 
signatures or some other systematic.  We 
will continue to plot these points in figures in this section but will not 
consider them in any modelling analysis or quotes of the significance of a 
signal.

\subsection{Cosmological Explanations For the Anti-Correlation Between 
QSOs and Galaxies}

An obvious physical effect, other than lensing, that 
could cause an anti-correlation between QSOs and 
galaxies is dust in galaxies obscuring background QSOs.  This would lead 
to a dearth of QSOs around galaxies.  Alternatively, if the 
anti-correlation {\it is} due to statistical lensing, we might be able 
to see the variation of the cross-correlation between galaxies and QSOs with the magnitude of 
the QSO sample.

\subsubsection{Is Intervening Dust the Cause of the Anti-Correlation 
Between QSOs and Galaxies?}
\label{dust}
\begin{figure}
\centering
\includegraphics[width=8cm,totalheight=8cm]{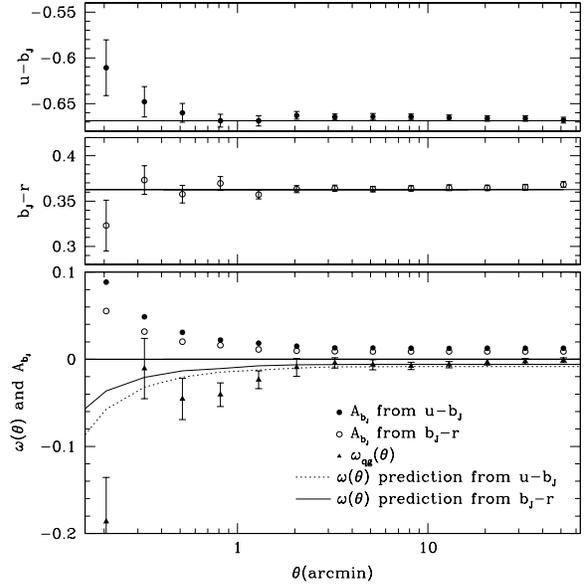}
\caption[\small{The Distribution of QSO Colours Around Galaxies.}]{Limits on the amount of `typical' dust that could account for the 
QSO-galaxy anti-correlation measured in Fig.~\ref{fig:qsoall5}.  The 
upper panels 
show the average colour of QSOs in bins centred on SDSS EDR galaxies (in the NGC 2QZ strip) and APM galaxies (in the SGC 2QZ strip). Error bars in these 
panels represent the standard deviation (1$\sigma$) in 1000 bootstrapped 
simulations with 
the same QSO positions as the 2QZ but scrambled colours.  The lower panels 
translate these 1$\sigma$ error bars into limits on absorption in the 
$b_{\rm J}$ band ($A_{b_{\rm J}}$).  The absorption limits 
are translated into limits on the QSO-galaxy cross-correlation using a simple model 
outlined in the text and displayed against the points and errors 
on $\omega_{qg}$ from Fig.~\ref{fig:qsoall5}, represented by triangles.}
\label{fig:dust5}
\end{figure}

We can use colour information to determine if intervening dust 
preferentially distributed around galaxies could remove QSOs from the 2QZ 
catalogue out to 10~arcmin ($\sim 1~\Mpch$).  Our method is similar to the 
correlation estimator of Equation~\ref{equation:corrfunc5} but instead of 
counting the average {\it number} of QSOs in differential annuli around 
galaxies, we 
work out the average {\it colour} of QSOs. Errors
are calculated using 1000 bootstrapped 
simulations with 
the same QSO positions as the 2QZ catalogue but scrambled colours.

Knowing 
the expected QSO colour on degree scales, we can calculate the allowed
colour excess, $E(B-V)$, in a given 
bin for both sets of measured 2QZ colours, $u-b_{\rm J}$ and $b_{\rm J}-r$.  Schlegel, Finkbeiner \& Davis (1998) provide tables to convert from the 
colour excess to the amount of absorption by dust. Thus we can constrain the amount of dust around galaxies along QSO lines of sight.

Our 1$\sigma$ limits on absorption can be converted into a 
limit on the correlation function using a simple model outlined in 
Boyle, Fong \& Shanks (1988).  Dust around galaxies will 
cause an absorption (which we calculate 
from our measured colour excesses) that will alter the magnitude 
limit of the QSO number counts close to galaxies
\begin{equation}
\omega + 1 = \frac{N(<m)_{Gal}}{N(<m)_{Field}} = \frac{N(<m 
-A_{b_{\rm J}})_{Field}}{N(<m)_{Field}}
\label{equation:absorption5}
\end{equation}
\noindent where $N(<m)$ is the integrated number counts, 
$A_{b_{\rm J}}$ is the absorption in the $b_{\rm J}$ band and the subscripts 
`Gal' and `Field' represent values near to galaxies and in the field, 
respectively.  Equation~\ref{equation:absorption5} can easily be
simplified if the number counts are represented by a power law but to be 
exact, we will simply use the full fitted form of the number counts of 
2QZ QSOs (see Fig.~\ref{fig:Nmag.eps}).

We now have a simple model that converts the error on our measurement 
of the average colour of QSOs near galaxies into a limit on the observed
anti-correlation due to dust.  In the upper two panels of Fig.~\ref{fig:dust5}, 
we show the average colour of QSOs around our galaxy samples with 
bootstrapped error bars, for 2QZ colours.  
A solid line marks the expected value from our scrambled bootstrap simulations.  
Except, perhaps, for the innermost bin, there is no significant deviation in the colour of QSOs from the expected value.  The reddening displayed in the innermost bin of the 
$u-b_{\rm J}$ colours corresponds to an increase to the blue in the 
$b_{\rm J}-r$ colours, 
suggesting it is a small-scale measurement artefact or a statistical fluctuation, rather than 
due to dust.  In the lower panel of Fig.~\ref{fig:dust5}, we translate the 
bootstrapped error bars into limits on absorption by dust around galaxies.  
The solid $u-b_{\rm J}$ and dashed $b_{\rm J}-r$ lines 
in the lower panel represent the 1$\sigma$ 
limit on the anti-correlation due to dust allowed by our colour limits.  
The anti-correlation between QSOs and galaxies measured in 
Fig.~\ref{fig:qsoall5} is plotted for comparison.  The 1$\sigma$ limits are insufficient to account for the anti-correlation 
between galaxies and QSOs, in fact, the $b_{\rm J}-r$ limits are marginally rejected,
at a 2$\sigma$ level, and (at this level of rejection) could only account for about 30~per~cent of the 
anti-correlation between QSOs and galaxies (out to 10~arcmin).

Schlegel, Finkbeiner \& Davis (1998)  base their absorption laws, from
which our absorption estimates are derived, on 
the difference in reddening between local, lightly-reddened standard 
stars and more distant reddened stars in the Milky Way \cite{Car89,Odo94}.  
The reddened stars are chosen to sample a wide range of interstellar 
environments \cite{Fit90} and Schlegel et al. demonstrate that their 
dust laws reproduce the reddening (as compared to the MgII 
index) of a sample of $\sim500$ elliptical galaxies that have broad sky 
coverage.  However, by design, the absorption law used by 
Schlegel et al. only applies to our Galaxy.  There is no 
reason to believe it is universal.  
In fact, though the Magellanic Clouds have been 
shown to have similar absorption laws to the Milky 
Way \cite{Koo82,Bou85}, other local ($z~\leqsim~0.03$) galaxies have
been shown to have `greyer' absorption laws \cite{Cal94,Kin94}, meaning 
that more dust absorption (approximately 25~per~cent in the $b_{\rm J}$
band) is expected for the same reddening. At the $2\sigma$ level, none of these alternative dust laws could provide enough absorption to explain the observed QSO-galaxy anti-correlation.

It is possible to construct dust models that would reconcile the lack of reddening of the QSO sample near galaxies with a paucity of QSOs around galaxies.  For instance, the UVX QSO selection of the 2QZ means that QSOs that were excessively reddened may be lost entirely from the 2QZ, especially if  the measured colours of 2QZ QSOs exhibit a large scatter around their intrinsic colours (which they don't, at least at $z < 2$ - see Croom et al. 2004).  Also, if there was a lot of dust close to some galaxies, which completely obscured QSOs without reddening them, then the clustering of galaxies alone might lead to the observed galaxy-QSO anti-correlation.  However, though various dust models can be constructed to explain a galaxy-QSO anti-correlation, it is very difficult to reconcile any dust model with the positive QSO-galaxy correlations seen in other samples \cite{Wil98,Gaz03}. 

Acknowledging the provisos outlined above, we tentatively proceed assuming that 
dust around galaxies cannot account for the anti-correlation between QSOs 
and galaxies---but can we find definitive evidence that gravitational 
lensing is responsible?

\subsubsection{Dependence of the Cross-Correlation Signal on Magnitude 
and Redshift}
\label{sec:magevolve}

Perhaps the key prediction of magnification bias is that an enhancement of QSOs is expected near foreground lenses when the slope of the QSO number-magnitude 
counts is greater than 0.4 and a deficit of QSOs around the same lenses 
when the slope is less than 0.4. The QSO number counts are
shown in Fig.~\ref{fig:Nmag.eps}.  The `knee' of 
the magnitude distribution, where the slope is 0.4, lies around 
$b_{\rm J}$ = 19.1 to $b_{\rm J}$ = 19.6.  A very simple model would 
predict (under the assumption that lensing samples QSOs up to about a 
magnitude fainter than the sample limit) a 
positive correlation between QSOs and galaxies up to a (QSO) $b_{\rm J}$ 
magnitude of around 18.1-18.6, no correlation between QSOs and galaxies in 
the range 18.1-18.6 to the knee of the magnitude counts, and an 
anti-correlation between QSOs and galaxies fainter than about $b_{\rm J}$ = 
19.6.

\begin{figure}
\centering
\includegraphics[width=8cm,totalheight=8cm]{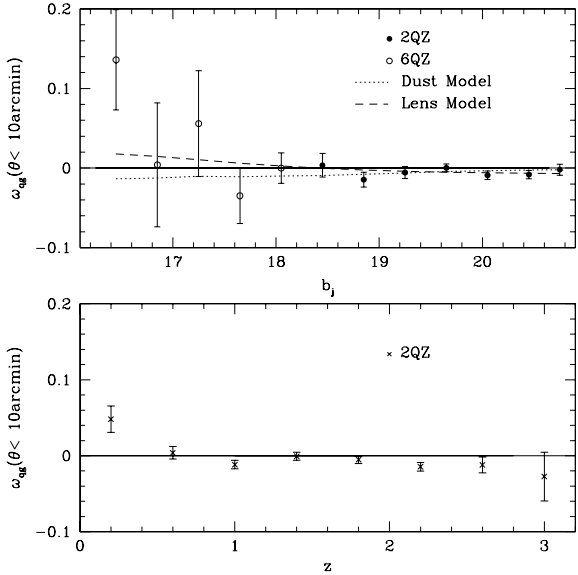}
\caption[\small{The Variation of $\omega_{gq}$ With Magnitude and 
Redshift.}]{The cross-correlation of 2QZ QSOs against SDSS
  EDR galaxies (in the NGC 2QZ  strip) and against APM galaxies
  (in the SGC 2QZ strip) and their dependence on $b_{\rm J}$ magnitude and 
redshift.  
The upper and lower panels display the cross-correlation signal measured out to 10~arcmin for subsamples of
  2QZ and 6QZ QSOs  in magnitude and redshift bins respectively. Error bars 
represent $1\sigma$ jackknife errors.  The `lensing' and `dust' models are described in the text.}
\label{fig:evolve5}
\end{figure}

At the time of writing, no author has yet shown the dependence of the 
cross-correlation signal with magnitude from a positive correlation at 
bright QSO magnitudes to an anti-correlation at faint magnitudes, although 
some authors have shown the transition from a positive correlation to zero 
correlation \cite{Wil98,Gaz03}.  This is mainly a problem of QSO sampling 
- as yet no single survey spans the QSO magnitude distribution in a manner 
that produces significant numbers of QSOs at both bright and faint 
magnitudes.  The combined area and depth of the 2QZ and 6QZ allows us
to trace the dependence of the cross-correlation between QSOs and
galaxies with magnitude across the knee of the QSO number-magnitude
counts for the first time.  In the upper panel of Fig.~\ref{fig:evolve5} we show the 
dependence of the QSO-galaxy cross-correlation signal (measured out to 
10~arcmin) in (differential) magnitude bins spanning the range $16.25 
< b_{\rm J} < 20.85$.  QSOs are taken from the 2QZ for $b_{\rm J} > 18.25$ 
and from the 6QZ for $b_{\rm J} < 18.25$. 
There is a loose trend in the data suggesting that the brightest QSOs are 
positively correlated with our galaxy samples (at the 2.2$\sigma$ level 
for $b_{\rm J} < 16.65$) and the faintest QSOs are anti-correlated with 
our galaxy samples (at the 2.3$\sigma$ level for $b_{\rm J} > 19.85$) and 
there is no significant result for the remainder of the magnitude range.  To test whether this trend allows us to favour lensing over dust in foreground galaxies as a cause of the QSO-galaxy cross-correlation signal, we have constructed two toy models.  The best-fitting models are displayed in Fig~\ref{fig:evolve5}.  The `dust' model is essentially as shown in Equation~\ref{equation:absorption5} with $A_{b_J}$, the absorption (on 10~arcmin scales to match the data in Fig.~\ref{fig:evolve5}) as the free parameter.  The relative numbers in a given magnitude bin are calculated from the full integrated counts displayed in Fig.~\ref{fig:Nmag.eps}.  The `lens' model assumes that $\omega_{qg} = K\left(2.5\alpha(m)-1\right)$ where $K$ is considered a constant (see Equation~\ref{equation:WilandIrwin} below) out to the 10~arcmin scale of interest.  There are actually two free parameters in this model, $K$ and $m$ where $m$ is the number of magnitudes fainter than the bin of interest to sample the slope, $\alpha$, of the integrated number counts.  The best dust model has $A_{b_J} = 0.005$ with a reduced $\chi^2$ value of 1.20 $\left(P(\chi^2)=0.29\right)$.  The best lens model has $K = 0.015$, $m = 2.1$ and a reduced $\chi^2$ of 1.04 $\left(P(\chi^2)=0.41\right)$.   Neither of these simple models can be ruled out by the data.  We cannot say with any confidence, then, that we 
have detected the full dependence of the cross-correlation signal with 
magnitude predicted by statistical lensing.  This is not necessarily 
surprising, as there are very few objects in the 6QZ and only 256 QSOs 
brighter than $b_{\rm J}=18.1$ that meet our usual redshift
and identification quality selection criteria.  Nevertheless, we do see a 
trend away from anti-correlations for brighter samples, in line with a lensing hypothesis.

In the lower panel of Fig.~\ref{fig:evolve5}, we show the variation of the 
cross-correlation of 2QZ QSOs against our combined galaxy samples with 
redshift in differential bins of $z = 0.4$.   A simple lensing model predicts no detectable 
trend in the cross-correlation signal with QSO redshift.  We might 
expect a stronger signal at larger redshifts, as we will, on 
average, sample fainter QSOs in these bins and thus observe the 
signature of changes in magnitude in the redshift distribution.  Indeed, 
we see a reasonably consistent anti-correlation for all redshift bins.  
The signal is slightly stronger at high redshifts but not significantly 
so.  The lowest redshift QSOs are significantly correlated with our galaxy 
samples, no doubt due to genuine clustering at these redshifts, justifying 
our cut of $z < 0.4$ in other analyses throughout this paper.

In this section, we have shown a significant 
anti-correlation between faint QSOs and galaxies, and that is not a systematic effect of the QSO selection.  We rule out the possibility that the majority of the signal is due to dust 
distributed around galaxies at the 2$\sigma$ level by comparing QSO colours in the field and close to galaxies, noting the caveat that models can be constructed where QSOs are obscured by dust without being reddened.  We have shown that the anti-correlation is, however, consistent with lensing predictions.  We will now 
model the anti-correlation assuming it is due to lensing and consider cosmological implications.

\section{Statistical Lensing Models}
\label{sec:model5}

In this section, we outline two lensing models we shall use in
describing the anti-correlation between QSOs and galaxies.  Both models compare the cross-correlation between QSOs and galaxies to the
auto-correlation of galaxies to determine galaxy bias with
respect to the mass traced by QSO light.  The first,
due to Williams \& Irwin (1998; see also Williams \& Frey 2003), uses a linear
biasing prescription to relate fluctuations in the foreground mass
distribution to QSO magnification.  Williams \& Irwin showed that
this simple model agrees well with more complicated simulations
\cite{Dol97,San97}.  The second model, due to Gazta\~naga (2003),
allows for scale-dependent bias, and more fully considers the
redshift distributions of the source and lens populations.

\subsection{Williams \& Irwin Model}
\label{sec:LBM}

The convergence of lensing matter,
$\kappa$, is defined as
\begin{equation}
\kappa(\theta) = \frac{\Sigma(D_l,\theta)}{\Sigma_{cr}(D_l,D_s)}
\label{equation:convergence}
\end{equation}
\noindent where $\Sigma(D_l,\theta)$ is the surface mass density of the lensing material.  The
critical mass surface density  is a function of the redshift of background
source QSOs ($z_s$) and of the foreground lensing matter ($z$), and is defined
\begin{equation}
\Sigma_{cr}(D_l,D_s) = \frac{c^2}{4\pi G}\frac{D_s}{D_lD_{ls}}
\label{equation:sigmacrit}
\end{equation}
\noindent for (angular diameter) lens distance $D_l$, source distance $D_s$ and
lens-source separation $D_{ls}$. We
take $z_s = 1.5$, the median redshift of the 2QZ, for the background
source redshift.  Although this seems a rather extreme approximation of the
actual distribution of QSOs, Bartelmann \& Schneider (2001) suggest it is fair. 

Williams \& Irwin model the lensing material as a smooth slab that extends
over the redshift range $z=0$ to
$z=z_{\rm{max}} = 0.3$, where
the extent of the redshift distribution is estimated using the magnitude-redshift
selection function of Baugh \& Efstathiou (1993), which provides a
good fit \cite{Mad96} to the Stromlo-APM Survey redshift
distribution \cite{Lov92a,Lov92b}.  We will also use the selection function of Baugh \& Efstathiou to model the galaxy redshift distribution.  Note that just over 95~per~cent of galaxies are
included out to $z_{\rm{max}} = 0.3$. 

Lensing arises due to angular fluctuations in the
projected matter density around the mean.  The mass fluctuations can be characterised by the density contrast $\delta(\theta)$, which measures the mass density at a given scale relative to the mean across the sky.  As the density across the entire sky {\it is} the mean, the distribution of density contrast is normalised $\left(\int P(\delta | \theta)\delta \rm{d}\delta = 1\right)$. The effective lensing convergence due to mass fluctuations are enhanced above the mean, yielding $\kappa_{\rm{eff}}(\theta) = \bar{\kappa}(\delta-1)$.   In this model, bias is scale-independent, meaning that galaxy fluctuations trace mass fluctuations as $\delta_{G}(\theta)-1 = b[\delta(\theta)-1]$. The surface density of a slab of matter of thickness $cdt$ at redshift $z$, is
$\Sigma = \rho_{crit}\Omega_0(1+z)^3cdt$, where $\rho_{crit}$ is the critical density of
the Universe. Thus

\begin{equation}
\kappa_{\rm{eff}}(\theta) = \frac{\bar{\kappa}}{b}(\delta_G-1) = \frac{3H_0^2c}{8\pi G}\frac{\Omega_0}{b}(\delta_{G}-1)\int_0^{z_{\rm{max}}}\frac{(1+z)^3\frac{dt}{dz}dz}{\Sigma_{cr}(z,z_s)} \label{equation:sigmaslab}
\end{equation}

The enhancement of QSO number density over
the mean around any galaxy is given, for QSO number-counts with
slope $\alpha$, by 

\begin{eqnarray}
\delta_{Q} = \mu^{2.5\alpha - 1} \approx \left(1+2 \bar{\kappa}\frac{(\delta_G-1)}{b}\right)(2.5\alpha -1)
\label{equation:deltaq}
\end{eqnarray}

\noindent where $\mu \approx 1+2\kappa_{\rm eff}$ is the lensing magnification in the weak regime (and we've performed a Taylor Expansion).   Now, the QSO-galaxy cross-correlation can be estimated by the enhancement in galaxy counts across the probability distribution of galaxy density contrasts

\begin{eqnarray}
\omega_{qg}(\theta) + 1 = \int P(\delta_{G} | \theta)\delta_{G}\delta_{Q}\rm{d}\delta_{G}
\end{eqnarray}

\noindent substituting Equation~\ref{equation:deltaq} into the previous equation and remembering $\int P(\delta | \theta)\delta \rm{d}\delta = 1$, one can rearrange to show

\begin{eqnarray}
\omega_{qg}(\theta) = (2.5\alpha - 1)\frac{2\bar{\kappa}}{b}\left[\int P(\delta_{G} | \theta)\delta_{G}\delta_{G}\rm{d}\delta_{G}\right] -1 
\end{eqnarray}

\noindent so, finally

\begin{eqnarray}
\omega_{qg}(\theta) = (2.5\alpha - 1)\frac{2\bar{\kappa}}{b}\omega_{gg}(\theta)
\label{equation:WilandIrwin}
\end{eqnarray}

\noindent relates $\omega_{gg}(\theta)$, to $\omega_{gg}(\theta)$ and $b$, the (scale-independent) bias parameter.  Other than $b$, cosmology is contained in the (dimensionless) mean convergence $\bar{\kappa}$ (Equation~\ref{equation:sigmaslab}).  The QSO number counts are 
included in the power-law slope, $\alpha$. 

\subsection{Gazta\~naga Model}

This model is reviewed in the Appendix, and described in detail by
Gazta\~naga (2003; see also Myers 2003). Gazta\~naga (2003)  has shown that 
a good power-law expression for scale-dependent biasing of galaxies 
is
\begin{eqnarray}
b(R) = b_{0.1}\left(\frac{0.1\Mpch}{R}\right)^{\gamma_b}
\label{biasmod}
\end{eqnarray}

\noindent Whilst Gazta\~naga  assumed that this bias corresponded to the ratio of the galaxy and matter correlation functions (see also Guim\~{a}raes 2001);
$b(r)=[\xi_{gg}(r)/\xi_{mm}(r)]^{0.5}$, we are in fact comparing 
the cross-correlation of QSOs
and galaxies with the  galaxy auto-correlation function, and so are 
defining a bias function via $b(r)=\xi_{gg}(r)/\xi_{gm}(r)$.
Whilst on large scales these two
definitions should be equal, one expects the halo of the central galaxy to affect
the shape of $\xi_{gm}(r)$ on small scales. Therefore we consider a
revised version of the Gazta\~naga model to take this into account.

Through observation of $\omega_{gg}$ and $\omega_{gq}$ we can constrain
the amplitude and slope of the bias correlation function.
  The 
two-dimensional projection of the galaxy correlation function is (consider Peebles 1980)
\begin{eqnarray}
\omega_{gg}(\theta) =
\sigma_{0.1}^2b_{0.1}b^*\theta^{1-\gamma_{gg}}A_{gg,0.1}
\label{equation:Agg}
\end{eqnarray}
\noindent whilst the galaxy-QSO cross-correlation is given by
\begin{eqnarray}
\omega_{gq}(\theta) =
\sigma_{0.1}^2b^*\theta^{1-\gamma_{gq}}A_{gq,0.1}
\label{equation:Aqg}
\end{eqnarray}
where $\sigma_{0.1}$, the amplitude of fluctuations in the matter correlation function (averaged over a sphere of radius 0.1$h^{-1}$Mpc) is defined in Equation~\ref{equation:sigsig},
$A_{gg,0.1}$ and $A_{gq,0.1}$ depend on the radial selection
functions of the galaxy and QSO samples and the lensing efficiency, and $b$ and $\gamma$ denote the slope and amplitude of a scale-dependent bias model.  In the particular case where the two bias definitions ($\left[\xi_{gg}/\xi_{gm}\right]$; $\left[\xi_{gg}/\xi_{mm}\right]^{0.5}$) are equivalent, $b*$ is equivalent to $b_{0.1}$ (see the Appendix for further details on these parameters, particularly $b*$).

The slope of the bias correlation function, $\gamma_b$, can be easily determined from $\gamma_{gg}$ and
$\gamma_{gq}$. The bias parameter, $b_{0.1}$, can
then be determined
\begin{eqnarray}
\label{equation:biascalc2}
b_{0.1} &=& \frac{\omega_{gg}(\theta)}{\omega_{gq}(\theta)}\frac{A_{gq,0.1}}{A_{gg,0.1}}\theta^{\gamma_b} 
\end{eqnarray}

In the case where the two bias definitions, above, agree exactly
(as implicitly assumed by Gazta\~naga),
$b^*=b_{0.1}$ and $\gamma^*=\gamma_b$. In this case, one can further
derive the slope of the mass correlation function, $\gamma$, and $\sigma_{0.1}$ can be determined via

\begin{eqnarray}
\sigma_{0.1}^2 &=&
\frac{\omega_{gq}(\theta)^2}{\omega_{gg}(\theta)}\frac{A_{gg,0.1}}{A_{gq,0.1}^2}\theta^{\gamma-1}.
\label{equation:biascalc}
\end{eqnarray}

\section{Results and Discussion}
\label{sec:discuss5}

\begin{figure}
\centering
\includegraphics[width=8cm,totalheight=8cm]{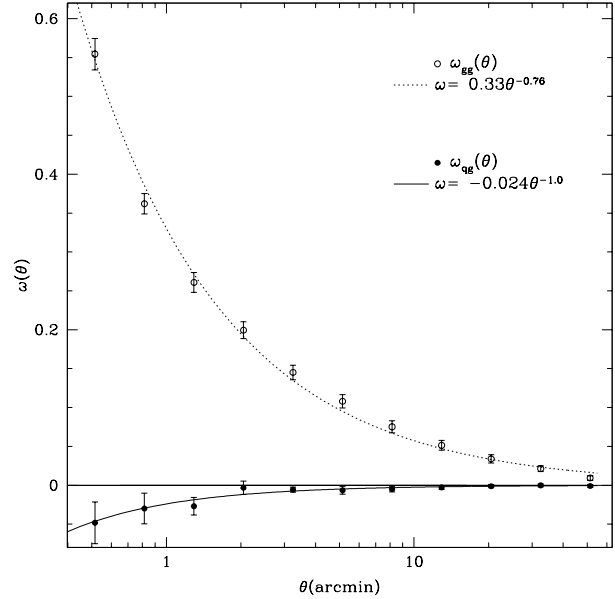}
\caption[\small{The Auto-correlation of Galaxies Compared to the
  Cross-Correlation of QSOs with Galaxies.}]{The 
galaxy-galaxy auto-correlation combined from
  SDSS EDR galaxies (in the NGC strip) and APM galaxies
  (in the SGC strip) is compared to the faint-end QSO-galaxy
  cross-correlation function.  Both correlation functions are fitted by 
  power laws, displayed by lines drawn through the respective data.  Error bars represent $1\sigma$ jackknife errors.}
\label{fig:gg5}
\end{figure}

In Fig.~\ref{fig:gg5} we plot the auto-correlation of galaxies combined 
for galaxies from the SDSS EDR in the NGC 2QZ strip and the APM Survey
in the SGC 2QZ strip.  Fig.~\ref{fig:gg5} also shows the
anti-correlation between QSOs and galaxies discussed throughout this
paper.  We have plotted the result for the QSO sample fainter than
$b_{\rm J}$ = 19.6, to ensure we are sampling QSOs from a region of the
number-magnitude counts fainter than the `knee' of the distribution
(see Section~\ref{sec:magevolve}).  We fit simple power-laws to the two
correlation functions, finding that
\begin{eqnarray}
\omega_{qg}(\theta) &=& -0.024\pm^{0.008}_{0.007}\theta^{-1.0\pm0.3}  \\
\omega_{gg}(\theta) &=& 0.330\pm^{0.015}_{0.014}\theta^{-0.76\pm0.04}
\label{equation:results5}
\end{eqnarray}
\noindent where $\theta$ is expressed in arcminutes, 1~arcminute being about
0.1$\Mpch$ at the median redshift of our galaxy data.  In both cases,
a simple $\chi^2$ fit suggests good power-law approximations
to the data, with a reduced $\chi^2$ value of 1.6 ($P(\chi^2)=0.11$) for the galaxy-galaxy correlation
fit and 1.4 $\left(P(\chi^2)=0.20\right)$ for the QSO-galaxy anti-correlation fit.
The result for the slope of the galaxy-galaxy auto-correlation is in excellent
agreement with recent small-scale measurements for galaxies with similar
limiting flux to the sample used here \cite{Con02}.

\subsection{Williams \& Irwin  Model}

\begin{table}
\centering
\begin{tabular}{cccc} \hline
         Model &       &
       {$\Omega_m = 0.3,\Omega_{\Lambda} = 0.7$} &
       {$\Omega_m = 1$} \\ \hline
        W\&I, 98 &
        $b_{0.1}$ & 0.13$\pm^{0.08}_{0.07}$    & 
0.32$\pm^{0.20}_{0.18}$              \\ \hline
        Gaz, 03 &        
        $b_{0.1}$ &  $ 0.052 \pm^{0.064}_{0.027}  $ &$ 0.142 \pm^{0.167}_{0.072}$\\ 
        &$b_{0.2}$ & $  0.061  \pm^{0.055}_{0.025}$&$ 0.166 \pm^{0.144}_{0.069}$\\ 
\hline
\end{tabular}
\caption[\small{The Galaxy Bias Parameter $b$ In Different Cosmologies.}]{\small{In the row labelled `W\&I, 98' we list $b$, evaluated at $\theta=1$~arcmin ($\sim0.1\Mpch$), as measured from
    Equation~\ref{equation:biasWilandIrwin5}, with $\Sigma_{cr}(z,z_s)$ calculated using 
    Equation~\ref{equation:sigmacrit} with $z_s = 1.5$ for the median QSO redshift. In the row labelled `Gaz, 03' we list $b$ evaluated
      at  $\theta=1$~arcmin, and $\theta=2$~arcmin
      ($\sim0.2\Mpch$), as measured from
      Equation~\ref{equation:biascalc2}.  Both estimates are shown for
      \LCDM and EdS cosmological models, assuming $H_0 = 70\kms\Mpc^{-1}$.}}
\label{table:beta}
\end{table}

Taking our measured values for 
the correlation functions, evaluated at $\theta=1$~arcmin ($\sim0.1\Mpch$),
and a faint end slope of $\alpha =0.29\pm0.03$,
Equation~\ref{equation:WilandIrwin} \cite{Wil98}  reduces to
\begin{eqnarray}
b_{0.1} = 7.56 \pm^{4.83}_{4.24} \bar{\kappa},
\label{equation:biasWilandIrwin5}
\end{eqnarray}
\noindent a function of the convergence, $\bar{\kappa}$. We
calculate this 
assuming either an Einstein-de-Sitter (henceforth EdS), or \LCDM
cosmology, and the
resulting values of $b$ are displayed in Table~\ref{table:beta}.  The model is somewhat dependent on the parameter $z_{\rm{max}}$ of Equation~\ref{equation:sigmaslab}.  For instance, increasing $z_{\rm{max}}$ from $0.3$ to $0.4$ increases the estimates of $b_{0.1}$ by 50~per~cent, with the errors scaling accordingly.  In our samples, 99.5~per~cent  of galaxies lie at $z < 0.4$.

Note
that, as the measured slopes of  $\omega_{gq}(\theta)$ and $\omega_{gg}(\theta)$
are not the same, a
scale-independent model of bias is only marginally accepted. It is relatively easy to extend
Equation~\ref{equation:WilandIrwin} to a simple model of
scale-dependent bias, obtaining

\begin{eqnarray}
b = b_{0.1}\left(\frac{0.1\Mpch}{r}\right)^{-0.24\pm0.30},
\label{equation:biasWilandIrwin6}
\end{eqnarray}

\noindent where $b_{0.1}$ is shown in Table~\ref{table:beta}
and $r$ is equivalent to $\theta$ expressed in $\Mpch$ at the mean
redshift of our galaxy data.

\subsection{Gazta\~naga Model}

Using the measured slopes $\gamma_{gg}$ and $\gamma_{gq}$, we find a
slope for the bias correlation function of
$\gamma_b = -0.24 \pm 0.30$.
We now use Equation~\ref{equation:biascalc2} to determine the galaxy
bias in both \LCDM and EdS cosmologies, with $H_0 = 70\kms\Mpc^{-1}$, for 0.1$\Mpch$ scales.
The results are displayed in Table~\ref{table:beta}. The
analysis is repeated measuring $b$ on $0.2\Mpch$ scales to
allow a direct comparison with the results of Gazta\~naga (2003).  The results only have a slight dependence on $H_0$.  Increasing $H_0$ to $100\kms\Mpc^{-1}$ increases the estimate of $b$ by 9~per~cent (for both EdS and \LCDM cosmologies, and for both $b_{0.1}$ and $b_{0.2}$). Decreasing $H_0$ to $50\kms\Mpc^{-1}$ decreases $b$ by either 8~per~cent (\LCDM) or 7~per~cent (EdS).

\begin{figure}
\centering
\includegraphics[width=8cm,totalheight=8cm]{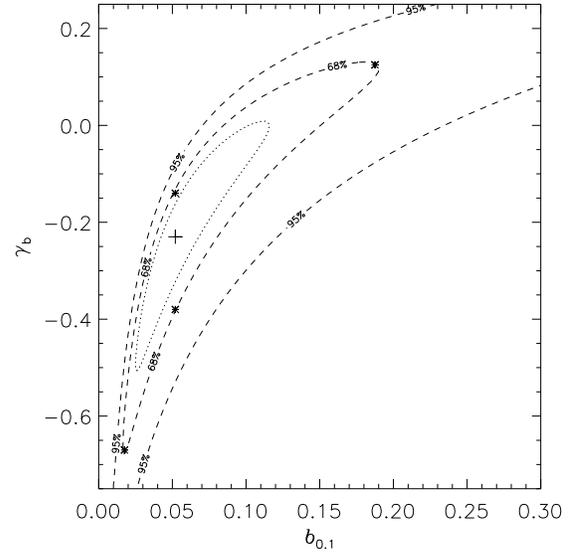}
\caption[\small{$b_{0.1}$ -- ${\gamma_b}$ likelihood
  contours}]{Likelihood contours in the $b_{0.1}$ -- ${\gamma_b}$ plane
for a fit to the QSO-galaxy
  cross-correlation function, $\omega_{gq}(\theta)$, assuming $\Lambda$CDM. Contours are
  plotted for $\chi^2$ values corresponding to a one-parameter confidence of 68 per cent (dotted contour), and
  two-parameter confidence of 68 and 95 per cent (dashed contours). The
  best fit model obtained (marked with a $+$) has $b_{0.1}=0.052$ and
  ${\gamma_b}=-0.23$. The stars mark the models plotted with dashed lines in Figure~\ref{fig:biascale}.
}
\label{fig:b01_gamma}
\end{figure}

Using the fit to $\omega_{gg}(\theta)$ from
Equation~\ref{equation:results5}, together with
Equation~\ref{equation:biascalc2}, we derive a model for
$\omega_{gq}(\theta)$ as a function of $b_{0.1}$ and ${\gamma_b}$. We
fit this model to the cross-correlation data by performing a maximum
likelihood analysis, determining the likelihood of each model by
calculating the $\chi^2$ value of each fit. Fig.~\ref{fig:b01_gamma}
shows likelihood contours in the $b_{0.1}$ -- ${\gamma_b}$ plane. There
is a clear degeneracy, with stronger scale-dependence of bias (i.e. more negative
$\gamma_b$) implying a lower value of $b_{0.1}$. 
The uncertainty in our determination of $b$ is somewhat larger than the
quoted error in Gazta\~naga (2003), and, as demonstrated by
Fig.~\ref{fig:b01_gamma}, is considerably skewed in
$b$-space. From Fig.~\ref{fig:b01_gamma} it is possible to reconcile the value of
$b_{0.1}$ obtained with the method of Gazta\~naga with the somewhat
higher value of $b_{0.1}$ obtained using the Williams \& Irwin model:
the latter assumes scale-invariant bias, or ${\gamma_b}=0$, which
corresponds to a higher value of $b_{0.1}\sim 0.12$  from the $b_{0.1}$
-- ${\gamma_b}$ degeneracy in the Gazta\~naga model fit.

\subsection{Discussion}

It is interesting that our
bias predictions, based on a detection of a galaxy-QSO {\it
  anti}-correlation, agree well with the predictions of an independent author \cite{Gaz03} who reported a {\it positive} galaxy-QSO correlation, working with bright QSOs and (mostly)
independent data.  
Both of the methods we have
used in this paper, which use quite
different lensing models to determine the amplitude of the galaxy bias
parameter, $b$, consistently agree that on scales of $0.1~\Mpch$,
galaxies are strongly anti-biased (i.e. $b < 1$) with a bias parameter
of $b \sim 0.1$.  Using a simple model \cite{Wil98} and assuming that
lensing matter can be represented by a uniform slab extending out to the
redshift where 95~per~cent of the surveyed galaxies are included, and
the lensed QSOs may be represented by a single population at $z = 1.5$, 
we find $b_{0.1} = 0.13 \pm^{0.08}_{0.07}$ for a \LCDM cosmology.  A more 
realistic model that traces
spherical fluctuations in the underlying lensing matter on $0.1\Mpch$
scales, across the redshift distributions of our QSO and galaxy
samples \cite{Gaz03}, yields $b_{0.1} = 0.04 \pm^{0.18}_{0.01}$.
Note that, whilst traditionally galaxy bias is measured from the ratio between the
galaxy auto-correlation and the mass auto-correlation, 
as we are comparing the cross-correlation of QSOs
and galaxies with the  galaxy auto-correlation function, we are instead
defining a bias function via $b(r)=\xi_{gg}(r)/\xi_{gm}(r)$. 

\begin{figure}
\centering
\includegraphics[width=8cm,totalheight=8cm]{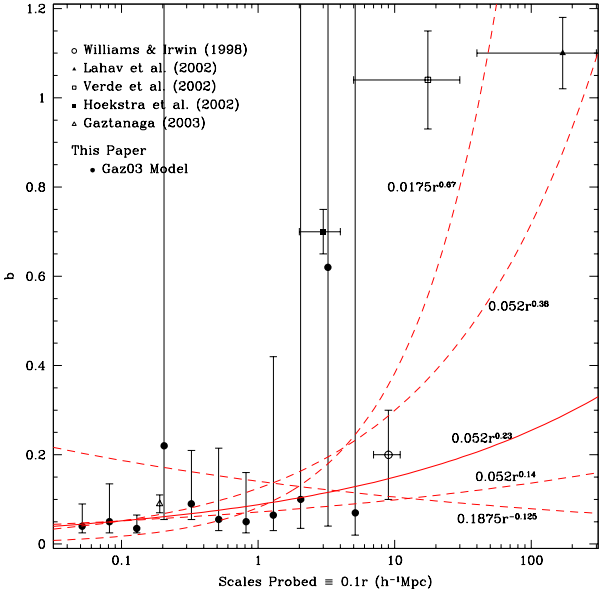}
\caption[\small{Recent Measurements of the Bias Parameter}]{A comparison of some recent measurements of the bias parameter
  with those made in this paper.  The filled circles are
  measurements of the bias parameter deduced from the ratio of the
  correlation functions in Fig.~\ref{fig:gg5} using
  Equation~\ref{equation:biascalc2} (i.e. the `Gazta\~naga Model').
  The angular bins of the correlation functions have been projected to
  the median redshift of the galaxy distribution ($z \sim 0.12$).  The
  solid line represents the best fitting bias model, with $b_{0.1}=0.052$ and
  ${\gamma_b}=-0.23$. The
  various dashed lines depict the error ranges taken from the 68\% likelihood
  contour; the models chosen are shown by stars in Figure~\ref{fig:b01_gamma}.  Each measurement displayed from this paper is calculated using a \LCDM cosmology.  Other points represent recent measurements of the bias parameter from the literature.}
\label{fig:biascale}
\end{figure}

A value of $b\sim0.1$  might seem at odds with observations of much higher values of  $b \sim 1$ on
Megaparsec scales \cite{Ver02,Lah02,Hoe02}, assuming a simple linearly-biased
 \LCDM mass distribution, suggesting more mass around galaxies
 than expected.   However, the slope of the scale-dependent bias parameter is found to be $\gamma_b = -0.24 \pm 0.3$. Although we cannot rule out a
linear bias parameter on scales of $0.1\Mpch$ from the slopes of our fitted
correlation functions, the fact that many other authors determine much
higher values of  $b \sim 1$ on
Megaparsec scales \cite{Ver02,Lah02,Hoe02} suggests that
bias might be a strong function of scale on $100\kpch$ scales.  In Fig.~\ref{fig:biascale} we plot the bias implied across a range of scales from measurements made in this paper and compare them to measurements made by other authors \cite{Wil98,Ver02,Lah02,Hoe02,Gaz03}.  The models discussed in this paper allow a scale-dependent model (with $b \sim r^{0.5}$)
 that is consistent both with our observations on $0.1\Mpch$ scales, as well as other 
measures of the galaxy bias parameter, with $b\sim1$, on larger scales
\cite{Lah02,Ver02}.   Whilst the models shown in
Fig.~\ref{fig:biascale} appear inconsistent with the measurements of
Hoekstra et al. (2002), it may be that the simple model of scale-dependent
bias assumed (equation~\ref{biasmod}) is not a good description of bias
on larger scales. Indeed, on very large scales, bias is expected to be
linear and hence independent of scale. On the other hand, Williams \& Irwin (1998) suggest galaxies
are highly anti-biased ($b \sim0.1$) even on 5-10$\Mpch$ scales, which would only be consistent with our observations if bias has no scale-dependence and is certainly inconsistent with the results of other authors \cite{Lah02,Ver02,Hoe02}.

To attempt to explain strong scale-dependence of bias, and, therefore, a 
 higher than expected  galaxy-mass cross-correlation signal on $100\kpch$ scales, we can consider
 halo occupation models of the galaxy distribution  \cite{Ber02,Jai03}.
  The form of the cross-correlation in a \LCDM universe
  can be determined from simulations, using halo
  occupation models to populate dark matter haloes with galaxies \cite{Ber02,Jai03},
  and taken in ratio with the
real-space galaxy auto-correlation function measured from the APM galaxy
catalogue \cite{Bau96}, such models typically
  yield a value of $b\sim0.6$ on $0.1\Mpch$ scales. Thus some level of
  anti-bias arises naturally in halo occupation models (albeit not as much as observed), going some way towards explaining the discrepancy. However, these models are complicated and
  depend strongly on galaxy type, and are thus not well constrained on
  the scales probed. Whilst a high QSO-galaxy cross-correlation signal places a constraint on such models, it may be possible to reconcile them with a better understanding of how galaxies populate haloes.

By comparing the
two definitions ($\left[\xi_{gg}/\xi_{gm}\right]$ and $\left[\xi_{gg}/\xi_{mm}\right]^{0.5}$) of $b(r)$ in a range of simulations, Berlind and
Weinberg (2002) demonstrated that whilst they are not in perfect
agreement in the scale-dependent bias regime ($0.1<r<4\,h^{-1}$\,Mpc),
the two definitions of $b(r)$ have a very similar shape over this
range. However, given the large anti-bias found, and strong
scale-dependence of bias implied, it is quite
likely that they {\it will in fact be considerably different}. Therefore,
care needs to be taken in interpreting the results of this analysis. In
particular, we have chosen not to follow the analysis of Gazta\~naga
(2003) in 
deriving the slope of the mass correlation function, $\gamma$, and
$\sigma_{0.1}$ under the assumption that the two bias definitions are
equal, as it is likely that this would lead to a large systematic
error, and hence misleading results.

Finally we should point out that in this work we have assumed the weak
lensing approximation, $\mu \approx 1+2\kappa$. Takada and Hamana
(2003) have shown that a non-linear magnification correction will enhance the
amplitude of the magnification correlation by around 10-25 per cent on
arcminute scales for the QSO-galaxy cross-correlation signal, and such a
correction would therefore boost our bias estimates by this
fraction. Whilst in the right direction, such a correction is too small
to explain the observed discrepancy in a \LCDM cosmology.  Note also that Table~\ref{table:beta} indicates  that an EdS cosmology with a value of $b_{0.1} \sim 1$
agrees with the data at the $2\sigma$ level.  Though an EdS framework
is consistent with the previous QSO lensing results of Croom \& Shanks
(1999) and Myers et al. (2003), it is inconsistent with the majority of recent
cosmological data.

\section{Conclusions}
\label{sec:summary5}

In this paper, we have studied the cross-correlation of galaxies and 
background QSOs.  Galaxies were drawn from the APM Survey in the region of the 2dF QSO Redshift Survey
Southern Galactic Cap strip and the Sloan Digital Sky Survey Early Data 
Release in the Northern Galactic Cap strip.   The QSO-galaxy
cross-correlation function $\omega_{gq}$ suggests that 2QZ QSOs and galaxies
are anti-correlated with significance ($2.8\sigma$) and strength
$-0.007$, over the angular range out to 10~arcmin (1$\Mpch$ at the
median redshift of our galaxy samples).  This result is unique in that
it is possibly the first
significant anti-correlation detected between QSOs and galaxies that is 
not subject to small sample sizes or problems in selecting the
populations. 

Our detection of a galaxy-QSO anti-correlation is consistent with
the predictions of statistical lensing theory.  When combined 
with the work of Gazta\~naga (2003),  a consistent picture
emerges that spans faint and bright QSO samples,
which, due to the changing QSO number-magnitude count slope, have very
different clustering properties, relative to the foreground galaxy
population.

We have also considered a number of other possible explanations for the anti-correlation between QSOs 
and galaxies.  Firstly, errors and the correlation estimator were
proven robust against Monte Carlo simulations.  Care was taken to demonstrate that there is no 
significant correlation (0.5$\sigma$ positive correlation) between
stars (which were selected within the 2dF Survey in 
the same way as the QSO sample) and our galaxy samples.  We are thus
confident that the anti-correlation between QSOs and galaxies is not a
systematic error.  The colours of QSOs around galaxies were used to place limits on
the effect dust (modelled with an absorption law appropriate to the
Milky Way) could have in producing
an anti-correlation of QSOs with galaxies.  Whilst our simple dust models
could account for at most
a third of the observed anti-correlation signal (at 2$\sigma$ significance),
we do not
rule out the possibility that scatter in QSO colours might preferentially discard reddened QSOs from the 2QZ, nor do we specifically rule out dust models that could obscure QSOs without reddening them, such as grey dust, or heavy concentrations of dust in some galaxies. No dust model, however, could easily explain the positive galaxy-QSO correlations found by Gazta\~naga (2003).

We find that, for a \LCDM cosmology, galaxies are highly anti-biased on small
scales.  We consider two models that use quite
different descriptions of the lensing matter and find they yield consistent
predictions for the strength of galaxy bias on $0.1\Mpch$ scales of $b
\sim 0.1$.  The inferred strength of this result is in agreement
with the work of Gazta\~naga (2003).  
The slope of the scale-dependent bias parameter is found to be
$\gamma_b = -0.24 \pm 0.3$. The fact that many other authors determine much
higher values of  $b \sim 1$ on 
Megaparsec scales \cite{Ver02,Lah02,Hoe02} suggests that
bias might be a strong function of scale. 

To  explain such strong scale-dependence of bias, we can consider
 halo occupation models of the galaxy distribution  \cite{Ber02,Jai03}.
Such models predict some level of
  anti-bias on $0.1\Mpch$ scales, typically
  yielding a value of $b\sim0.6$ \cite{Ber02}. However, these models are complicated and
  depend strongly on galaxy type, and are thus not well constrained on
  the scales probed. Therefore, it may be possible to reconcile them
  with the observations reported here through 
  a better understanding of how galaxies populate haloes.

An alternative interpretation of these results is that they indicate
that there is significantly
more mass present, at least on the $100\kpch$ scales probed, than predicted by
\LCDM. Myers et al. (2003) recently detected a strong anti-correlation between
the same faint 2QZ QSOs and groups of galaxies.
By applying models of gravitational 
lensing by simple haloes they used this anti-correlation signal to
determine the mass of these lensing galaxy groups, concluding that the observed anti-correlation 
favours considerably more mass in groups of
galaxies than accounted for in a universe with density parameter 
$\Omega_m = 0.3$. It is hard to account for the statistical lensing
properties either of the 
galaxies presented here, or the groups of galaxies from Myers et al., in
a low mass ($\Omega_m \sim 0.3$) universe with scale-independent bias. 

 \section*{Acknowledgements} 
The 2dF QSO Redshift Survey was based on observations made with the
Anglo-Australian Telescope and the UK Schmidt Telescope, and we would
like to thank our colleagues on the 2dF Galaxy Redshift Survey team and
all the staff at the AAT that have helped to make this survey
possible. ADM ackowledges the support of a PPARC studentship. PJO acknowledges the support of a PPARC Fellowship.
This work was partially supported by the `SISCO' European Community Research
and Training Network. 

Funding for the creation and distribution of the SDSS Archive has been provided by the Alfred P. Sloan Foundation, the Participating Institutions, the National Aeronautics and
Space Administration, the National Science Foundation, the
U.S. Department of Energy, the Japanese Monbukagakusho, and the Max
Planck Society. The SDSS Web site is {\tt http://www.sdss.org/}. 

The SDSS is managed by the Astrophysical Research Consortium (ARC) for the Participating Institutions. The Participating Institutions are The University of Chicago, Fermilab, the Institute for Advanced Study, the Japan Participation Group, The Johns Hopkins University, Los Alamos National Laboratory, the Max-Planck-Institute for Astronomy (MPIA), the Max-Planck-Institute for Astrophysics (MPA), New Mexico State University, University of Pittsburgh, Princeton University, the United States Naval Observatory, and the University of Washington.

\appendix

\section{Gazta\~naga Model}

The model reviewed in this appendix is described in detail by
Gazta\~naga (2003; see also Myers 2003).

Groth \& Peebles (1977) and Phillipps et al. (1978) proposed the `$\epsilon$' 
form of the (three-dimensional) correlation function to describe the evolution of clustering with redshift
\begin{eqnarray}
\xi(r,z) = \left(\frac{r_0}{r}\right)^{\gamma}(1+z)^{-(3+\epsilon)}
\end{eqnarray}
\noindent For the purposes of this
model we shall assume that $\epsilon = 0$, corresponding to the
stable clustering regime, in reasonable agreement with both
current observational evidence and  theoretical
considerations \cite{Wilson03,Car00,Ham91,Pea94} for the small-scale correlations 
of lensing matter we consider,  at the low redshifts ($z \sim 0.15$) where
our galaxy samples reside. 

As at the median depth of our galaxy samples ($z\sim0.15$), 1~arcmin corresponds to
0.1$\Mpch$, we choose to measure fluctuations on this scale. The
volume-averaged integral of the mass correlation function, $\xi$, over a sphere of radius $R$ is \cite{Pee80}
\begin{eqnarray}
\bar{\xi}(R) = \sigma_{0.1}^2\left(\frac{0.1\Mpch}{R}\right)^\gamma
\label{equation:corravap}
\end{eqnarray}
\noindent with
\begin{eqnarray}
\sigma_{0.1}^2 = 
\frac{72}{(3-\gamma)(4-\gamma)(6-\gamma)2^\gamma}\left(\frac{r_0}{0.1\Mpch}\right)^\gamma .
\label{equation:sigsig}
\end{eqnarray}

Gazta\~naga (2003)  has shown that 
a good power-law expression for scale-dependent biasing of galaxies 
is
\begin{eqnarray}
b(R) = b_{0.1}\left(\frac{0.1\Mpch}{R}\right)^{\gamma_b}
\end{eqnarray}

Whilst Gazta\~naga  assumed that this bias corresponded to the ratio of the galaxy and matter correlation functions;
$b(r)=[\xi_{gg}(r)/\xi_{mm}(r)]^{0.5}$, we are in fact comparing 
the cross-correlation of QSOs
and galaxies with the  galaxy auto-correlation function, and so are 
defining a bias function via $b(r)=\xi_{gg}(r)/\xi_{gm}(r)$.
Whilst on large scales these two
definitions should be equal, one expects the halo of the central galaxy to affect
the shape of $\xi_{gm}(r)$ on small scales. Therefore we consider a
revised version of the Gazta\~naga model to take this into account.

Firstly we define the volume-averaged galaxy-mass cross-correlation,
$\bar{\xi}_{gm}(R)$ via,
\begin{eqnarray}
\bar{\xi}_{gm}(R) = b^* \left(\frac{0.1\Mpch}{R}\right)^{\gamma^*}
\times \bar{\xi}(R).
\end{eqnarray}
\noindent In the case where the two definitions of bias given above agree exactly,
$b^*=b_{0.1}$ and $\gamma^*=\gamma_b$.

The galaxy-averaged correlation function can 
thus be expressed in terms of 
the mass-averaged function of Equation~\ref{equation:corravap} as
\begin{eqnarray}
\bar{\xi}_g(R) = 
b_{0.1}b^*\sigma_{0.1}^2\left(\frac{0.1\Mpch}{R}\right)^{\gamma+\gamma^*+\gamma_b}.
\end{eqnarray}

The 
two-dimensional projection of the galaxy correlation function is (again consider Peebles 1980)
\begin{eqnarray}
\omega_{gg}(\theta) =
\sigma_{0.1}^2b_{0.1}b^*\theta^{1-\gamma_{gg}}A_{gg,0.1}
\end{eqnarray}
where
\begin{eqnarray}
A_{gg,0.1}=B_{0.1}(\gamma_{gg})\int d\chi\left(\frac{dN_g}{d\chi}\right)^2\chi^{1-\gamma_{gg}}(1+z)^{-(3+\epsilon)}
\end{eqnarray}
with
\begin{eqnarray}
B_{a}(\gamma) = 
\frac{\gamma(3-\gamma)(4-\gamma)(6-\gamma)2^\gamma}{a^{-\gamma}72}\frac{\Gamma(1/2)\Gamma(\gamma/2-1/2)}{\Gamma(\gamma/2)}
\end{eqnarray}
\noindent and $\gamma_{gg}=\gamma+\gamma^*+\gamma_b$. Note that $dN_g/d\chi$ is the galaxy radial selection function as a function of comoving distance, $\chi$, which is easily derived from the normalised redshift distribution of the galaxies \cite{Gaz03}.  We use the selection function of Baugh and Efstathiou (1993) to model the redshift distribution of our galaxies.

In the weak 
lensing approximation, we can
relate the fluctuations in projected QSO numbers due to lensing to the convergence via \cite{Bar95,Bar01,Men02}
\begin{eqnarray}
\delta_{\mu}(\theta) &=& (2.5\alpha-1)\delta\mu = 2(2.5\alpha-1)\kappa_{\rm{eff}}(\theta)\\
&=& \int d\chi \varepsilon(\chi) \delta(\theta,\chi).
\end{eqnarray}
\noindent where $\delta(\theta,\chi)$ is the density contrast of the underlying
lensing matter. The lensing efficiency, $\varepsilon(\chi)$, is defined \cite{Ber97} as
\begin{eqnarray}
\varepsilon(\chi) = (2.5\alpha-1)\frac{3{H_0}^2\Omega_m}{c^2}(1+z)\chi\int^\infty_\chi d\chi'\frac{\chi'-\chi}{\chi'}\frac{dN_q}{d\chi'}
\end{eqnarray}
\noindent where $dN_q/d\chi$ is the QSO selection function in comoving
coordinates, which we derive from the redshift distribution of (quality `11', $z > 0.4$) QSOs in the 2QZ sample.

As outlined in Gazta\~naga (2003), these equations allow us to derive
the galaxy-QSO cross-correlation.  
\begin{eqnarray}
\omega_{gq}(\theta) =
\sigma_{0.1}^2b^*\theta^{1-\gamma_{gq}}A_{gq,0.1}
\end{eqnarray}
where
\begin{eqnarray}
A_{gq,0.1}=B_{0.1}(\gamma_{gq})\int d\chi\left(\frac{dN_g}{d\chi}\right)\varepsilon(\chi)\chi^{1-\gamma_{gq}}(1+z)^{-(3+\epsilon)}
\end{eqnarray}
\noindent where $\gamma_{gq}=\gamma+\gamma^*$. $\varepsilon(\chi)$ is typically less than
1~per~cent of $\frac{dN_G}{d\chi}$, which often leads to the
prediction that $\omega_{gq}$ should only be observed to be a tiny fraction of
$\omega_{gg}$ \cite{Dol97}.

The slope of the bias correlation function, $\gamma_b$, can be easily determined from $\gamma_{gg}$ and
$\gamma_{gq}$. The bias parameter, $b_{0.1}$ can be determined via

\begin{eqnarray}
\label{equation:biascalc2a}
b_{0.1} &=& \frac{\omega_{gg}(\theta)}{\omega_{gq}(\theta)}\frac{A_{gq,0.1}}{A_{gg,0.1}}\theta^{\gamma_b} 
\end{eqnarray}

If the two definitions of bias given above agree exactly,
$b^*=b_{0.1}$ and $\gamma^*=\gamma_b$, and one could further
derive the slope of the mass correlation function, $\gamma$, and $\sigma_{0.1}$ via

\begin{eqnarray}
\sigma_{0.1}^2 &=&
\frac{\omega_{gq}(\theta)^2}{\omega_{gg}(\theta)}\frac{A_{gg,0.1}}{A_{gq,0.1}^2}\theta^{\gamma-1}.
\label{equation:biascalca}
\end{eqnarray}

\end{document}